\documentclass[epj]{svjour}

\usepackage{graphics}
\usepackage{amsmath}
\usepackage{times}
\begin{document}
\titlerunning{Image windowing for DDM}
\title{Correcting artifacts from finite image size in Differential Dynamic Microscopy}
\author{Fabio Giavazzi\inst{1}, Paolo Edera \inst{1}, Peter J.~Lu \inst{2} \and Roberto Cerbino\inst{1}
}                     
%
%
\institute{Dipartimento di Biotecnologie Mediche e Medicina Traslazionale, Universit\`a degli Studi di Milano, via F.lli Cervi 93, 20090 Segrate, Italy \and Department of Physics and SEAS, Harvard University, Cambridge, Massachusetts 02138 USA}
\date{Received: date / Revised version: date}
%
\abstract{
Differential Dynamic Microscopy (DDM) analyzes traditional real-space microscope images to extract information on sample dynamics in a way akin to light scattering, by decomposing each image in a sequence into Fourier modes, and evaluating their time correlation properties. DDM has been applied in a number of soft-matter and colloidal systems. However, objects observed to move out of the microscope's captured field of view, intersecting the edges of the acquired images, can introduce spurious but significant errors in the subsequent analysis. Here we show that application of a spatial windowing filter to images in a sequence before they enter the standard DDM analysis can reduce these artifacts substantially. Moreover, windowing can increase significantly the accessible range of wave vectors probed by DDM, and may further yield unexpected information, such as the size polydispersity of a colloidal suspension.
} 
\maketitle
\section{Introduction}
\label{intro}
Differential Dynamic Microscopy (DDM) uses Fourier analysis of microscope image sequences to characterize the structure and dynamics of a wide variety of physical and biological systems, including dilute isotropic \cite{Cerbino:2008if,Bayles:2016_darkDDM} and anisotropic \cite{Reufer2012,Giavazzi:2016_jpcm} colloidal particles, dense colloidal suspensions~\cite{Lu:2012oa,Laurati2016,Kodger:2017mw}, molecular \cite{Giavazzi2016_epje} and complex \cite{savo2016,Giavazzi:2014fi} fluids, motile microorganisms \cite{Lu:2012oa,Wilson:2011wa,Martinez:2012ya}, and sub-cellular structures \cite{Drechsler2017,Feriani2017}. This broad adoption of DDM stems from its numerous advantages~\cite{Giavazzi:2014sj}, including simple implementation with ordinary microscopy, no need for custom instrumentation, insensitivity to normal amounts of dirt or multiple scattering, and an ability to focus on regions of interest in images collected with a variety of image-contrast mechanisms: bright field \cite{Cerbino:2008if}, dark-field \cite{Bayles:2016_darkDDM}, phase contrast \cite{Wilson:2011wa}, wide field fluorescence \cite{He:2012bx}, polarized \cite{Giavazzi:2016_jpcm,Giavazzi:2014fi}, 
differential interference contrast \cite{Drechsler2017}, light sheet \cite{Wulstein:16} and confocal microscopy (ConDDM) \cite{Lu:2012oa,Laurati2016,Kodger:2017mw}.

Theoretically, DDM probes a range of wave-vectors $q$ that are determined by two factors: the lower bound $q_{\mathrm{min},\mathrm{th}}=2\pi/L$ is constrained by the image size $L$, while the upper bound $q_{\mathrm{max},\mathrm{th}}=\pi/a$ is controlled by the pixel size $a$. In real experiments, the practical range $\left[ q_{\mathrm{min}},q_{\mathrm{max}}\right]$ for which the statics and the dynamics can be measured reliably is often more limited. Limitations can arise from both the statics (e.g. the signal to noise ratio is too low) and the dynamics (e.g. the observation time window is too short to adequately sample the dynamics associated with the slowest modes, or the temporal resolution is too poor to capture the faster dynamics, typically associated with the smaller length scales). Other relevant practical limitations may result from mechanical drifts, vibrations or advective/convective flows driven by thermal inhomogeneities or pressure imbalance.

An additional limitation constraining the range of probed wavevectors arises from the fact that, in any sequence of images with finite size, particles crossing the edge of the image boundary will be imaged only partially. Thus, the images contain particles with straight, sharp edges that, as is well-known in signal processing theory~\cite{Priemer1990}, create significant artifacts in the Fourier spectrum. This effect is particularly pronounced in systems with limited spatial bandwidth, as is common in microscope images due to the resolution constraints imposed by the diffraction limit. Although, thus far, this problem has been given little attention, it nonetheless leads to spurious artifacts in the Fourier transforms of the images, thereby potentially affecting both the effective $q$ range that can be probed with DDM, though the specific effects have not yet been established.

In this paper, we combine theory and experiments to show that the partial imaging of particles at the boundary, inevitable for all images of finite size, introduces significant artifacts, namely a spurious, nearly-$q$-independent secondary decay in the DDM image structure functions. This decay, present in principle for all $q$, dominates the dynamics at the largest $q$ values, where the signal associated with particle dynamics vanishes due to minima of the particle form factor $P(q)$. We mitigate this artifact with a simple preprocessing step: spatial windowing (apodization) of the images, which does not increase substantially computational complexity, yet increases significantly $q_{max}$. The expansion of the accessible range of $q$ values not only improves the accuracy of DDM in general, but also opens up new analyses in specific cases; for example, we show how windowing may enable the estimation of size polydispersity in a colloidal suspension using a method common in Dynamic Light Scattering (DLS) experiments \cite{Pusey1984,schope2007}. 

\section{Boundary effects in dynamic microscopy}
\label{sec:1}

A detailed description of the image processing algorithm on which DDM is based can be found in Refs. \cite{Croccolo:2006et,Cerbino:2008if,Giavazzi:2009xd,Giavazzi:2014sj}. In brief, a sequence of $N$ digital images
$I(\mathbf{x},t)$ is acquired, where $\mathbf{x}=a_{0}\,(n_{x},n_{y})$
and $t=n\Delta t_{0}$. Here $a_{0}$ is the effective pixel size
(the physical pixel size divided by the objective magnification),
$n_{x},n_{y}$ are integer numbers comprised between $1$ and the image size $M$ (assumed to be the same for both dimensions) and $\Delta t_{0}$ is the time interval between two consecutive images. The key quantity from which the dynamical information is extracted is the so-called image structure function $D(\mathbf{q},\Delta t)$, that is calculated as
\begin{equation}
D(\mathbf{q},\Delta t)=\left\langle \left|\mathit{FFT}\left[I(\mathbf{x},t_{0}+\Delta t)-I(\mathbf{x},t_{0})\right]\right|^{2}\right\rangle \label{eq:struf-1}
\end{equation}
where $\mathit{FFT}$ indicates the the Fast Fourier Transform operation and $\mathbf{q}=q_{0}\,(m_{x},m_{y})$, with $m_{x,}m_{y}$ integers comprised
between $-\left(\frac{M}{2}-1\right)$ and $\frac{M}{2}$. $q_{0}=\frac{2\pi}{Ma_{0}}$. The expectation value $\langle \cdot \rangle$ is taken over time and, possibly, over different replicas of the same experiment. 

\begin{figure*}
\resizebox{2\columnwidth}{!}{%
  \includegraphics{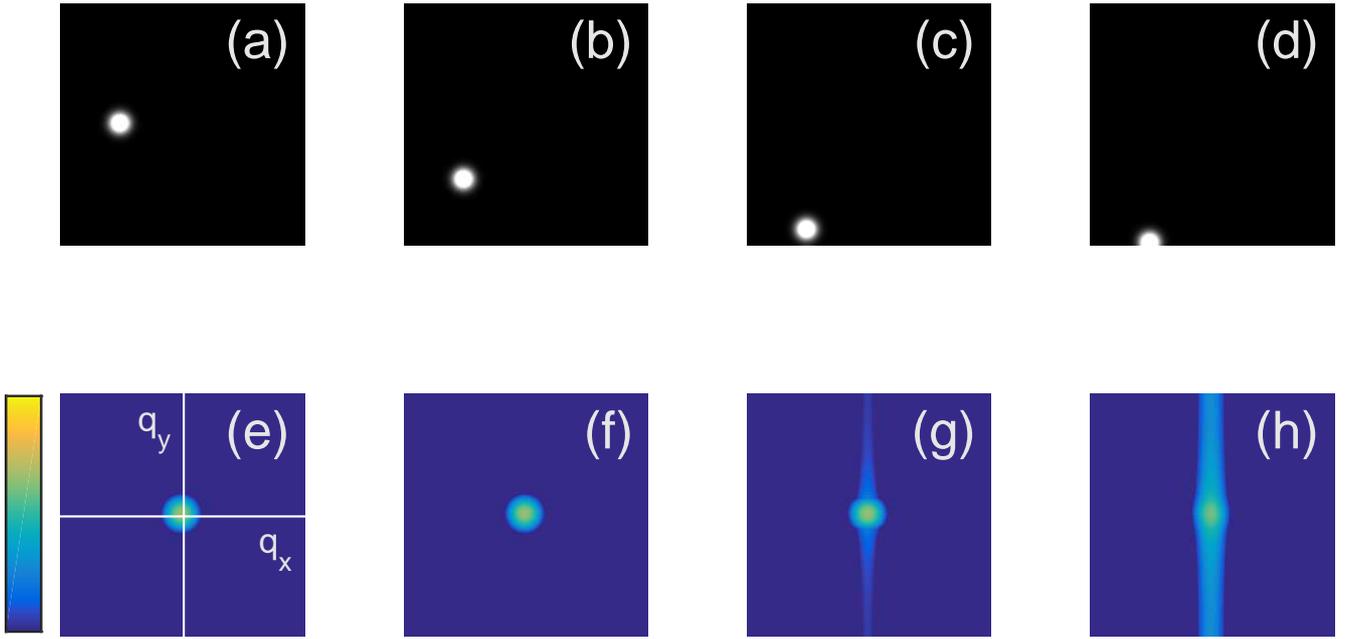}
}
\caption{Spectral leakage in DDM experiments. Particles crossing image boundaries (top row) excite high-$q$ wave-vectors in the reconstructed Fourier spectrum (bottom row) along the direction perpendicular to the image boundary. Consequently, the characteristic dynamics associated with these events show strong effects at large $q$. All images are collected with the same settings, and Fourier spectra are represented on a logarithmic scale with the same color code.}
\label{fig:1}       
\end{figure*}
For a linear space-invariant imaging process, the image structure function takes the form \cite{Giavazzi:2009xd}
\begin{equation}
D(\mathbf{q},\Delta t)=A(\mathbf{q})\left[1-f(\mathbf{q},\Delta t)\right]+B(\mathbf{q})
\label{cumpa}
\end{equation}
where $A(\mathbf{q})$ is an amplitude term that depends on the spatial intensity correlations present in the images and $B(\mathbf{q})$ accounts for the noise of the detection chain. The function $f(\mathbf{q},\Delta t)=f_{2D}(\mathbf{q},\Delta t)f_{z}(\mathbf{q},\Delta t)$ is defined in terms of a transverse part $f_{2D}$ encoding dynamics in the image plane and an axial contribution $f_{z}$, which accounts for dynamics in the axial direction. In most cases of interest, such as for instance when the axial dynamics can be neglected or when small wave-vectors are of interest, $f(\mathbf{q},\Delta t)$ coincides with
the normalized intermediate scattering function probed by DLS \cite{Berne:2000ye,Giavazzi:2014sj}.

Starting from Eq. \ref{cumpa}, the usual strategy in DDM experiments is based on 
\begin{enumerate}
\item assuming a suitable functional form describing the time dependence of $f(\mathbf{q},\Delta t)$
\item fitting the image structure function $D(\mathbf{q},\Delta t)$ to estimate the $q$-dependent parameters describing the relaxation of the different Fourier modes
\item collecting together the results obtained at different $q$ to extract the relevant quantity characterizing the dynamics and the statics of the sample.
\end{enumerate}

For example, for a dispersion of dilute, non-interacting Brownian particles, the expected intermediate scattering function is $f(q,\Delta t)=\exp\left(-\Gamma(q)\Delta t\right)$. The fitting procedure provides an estimate of $\Gamma(q)$, whose expected scaling with $q$ is $\Gamma(q)=D_{t} q^2$, where $D_{t}$ is the translational diffusion coefficient of the particles. The best estimate for $D_{t}$ is then obtained by a fit of $\Gamma(q)$. In this particular case, no structural correlations are expected, which means that the estimate for $A(\mathbf{q})$ provided by the fitting procedure provides information about the form factor $P(\mathbf{q})$ of the particles and the transfer function $T(\mathbf{q})$ of the optical setup \cite{Giavazzi:2009xd}. In other cases, additional information about the structural correlations within the sample can be extracted \cite{Lu:2012oa,savo2016,Giavazzi:2014fi}.

Further insight can be obtained by making explicit the relationship between the sampled intensity $I(\mathbf{x},t)$ on the detector and the actual intensity $i(\mathbf{x},t)$ in the image plane as
\begin{equation}
I(\mathbf{x},t)=W_0(\mathbf{x})\left[i(\mathbf{x},t) +b(\mathbf{x},t)\right], \label{LINSPI2}
\end{equation}
which is helpful to account for finite sampling effects. Here, $W_0(\mathbf{x})$ is a window function that takes value $1$ within the image boundaries and $0$ outside and $b$ is a detection noise term that we assume to be delta-correlated both in space and time.

In the following, we will focus on the case of a collection of $N_p$ identical particles, whose positions are labeled by the coordinates $(\mathbf{x}_{n},z_{n})_{n=1,2,..,N_p}$. For a linear, space-invariant imaging process \cite{Giavazzi:2009xd}, we obtain
\begin{equation}
i(\mathbf{x},t)=i_0+\sum_{n}\psi\left(\mathbf{x}-\mathbf{x}_{n}(t),z_{n}(t)\right)\label{LINSPI}
\end{equation}
where $i_0$ is the average intensity in the absence of the particles and $\psi$ represents the
intensity distribution associated with a single particle. In general, $\psi$ is the result of the 2D convolution of the spatial distribution of the relevant optical parameter within the particle (\textit{e.g.} refractive index in the case of bright-field or dye density in the case of florescence microscopy) with the three-dimensional point-spread function of the microscope \cite{Giavazzi:2009xd}.
By introducing the spatial 2D Fourier transform of the function $g(\mathbf{x})$:
\begin{equation}
\hat{g}(\mathbf{q})=\int_{-\infty}^{+\infty}dx\int_{-\infty}^{+\infty}dyg(\mathbf{x})e^{-j\mathbf{q}\cdot\mathbf{x}}. \label{four_trans}
\end{equation}

we obtain after some manipulation the following expressions for the intermediate scattering function $f(\mathbf{q},\Delta t)$ and the amplitude $A(\mathbf{q})$:

\begin{equation}
A(\mathbf{q})=2\tilde{N}_{p} P(\mathbf{q})
\end{equation}

and \begin{equation}
f(\mathbf{q},\Delta t)=\frac{|\hat{W_0}(\mathbf{q})|^{2}\ast\left[f_{\infty}(\mathbf{q},\Delta t)P_{\infty}(\mathbf{q})\right]}{P(\mathbf{q})}
\label{gtilde}
\end{equation}
where $\tilde{N_p}$ is the average number of particles within the image and where we have defined the form factor

\begin{equation}
P(\mathbf{q})=|\hat{W_0}(\mathbf{q})|^{2}\ast P_{\infty}(\mathbf{q})
\label{ptilde}
\end{equation}

and its limit for infinitely large samples
\begin{equation}
P_{\infty}(\mathbf{q})=\langle |\hat{\psi}\left(\mathbf{q},z\right)|^{2}\rangle.
\end{equation}

The noise term $B(\mathbf{q})$ is expected to be $q$ independent and proportional to $\langle b^2 \rangle$.
These equations describe how the statical and dynamical properties of particles, when reconstructed from the FFT analysis of the images, are affected by the presence of the boundaries and may differ from the ones calculated for an infinitely extended image, i.e. when $\hat{W_0}(\mathbf{q}) \simeq \delta(\mathbf{q})$. In fact, only in the latter case the intermediate scattering function is given by $f(\mathbf{q},\Delta t)=f_{\infty}(\mathbf{q},\Delta t)$. In all other cases, a mixing between different Fourier components occurs, which for the static amplitude is known as \textit{spectral leakage} in the signal processing literature \cite{Harris1978}.


\begin{figure*}
\centering
\resizebox{1.75\columnwidth}{!}{%
  \includegraphics{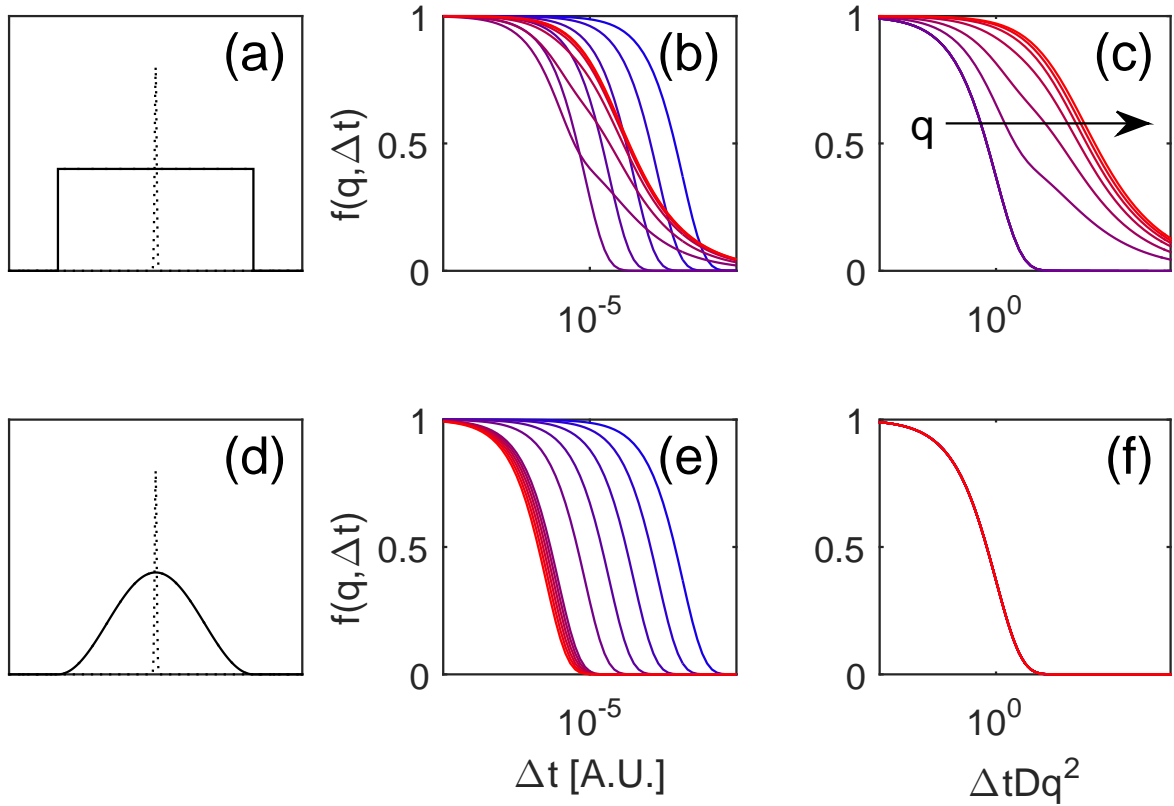}
}
\caption{Results obtained by numerical integration of Eq. \ref{gtilde} in the case of Brownian particles with diffusion coefficient $D_{t}=1$ and Gaussian effective shape with standard deviation $\sigma_P=0.05$ (dotted profile in panels (a) and (d)). The image size is assumed to be $L=1$. In the absence of explicit windowing, the window function coincides with the characteristic function of the image area (continous line in panel (a)). In panel (b) the corresponding normalized intermediate scattering functions for different $q$s in the range $0.1<q<10$ are shown. For large $q$, the curves converge to a $q$-independent decay. Such deviation from the expected exponential behavior $f(q,\Delta t)=\exp(-D_{t}q^2\Delta t)$ is made evident in panel (c), where the same curves are plotted as a function of the rescaled time $D_{t}q^2\Delta t$. The application of a smooth window function (continous curve in panel (d), see also Eq. \ref{wind_funct}) substantially reduces the spurious affects on the dynamics, as shown in panels (e-f) where the corresponding normalized intermediate scattering functions are shown for the same $q$ values considered in panels (b-c).}
\label{fig:df}       
\end{figure*}

To have a physical intuition of why spectral leakage also affects the dynamics, one can consider Fig. \ref{fig:1}, where simulated real-space images of a spherical particle in different positions (a-d) are compared with the corresponding $FFT$ spectra (e-h). As far as the particle is well within the image area, the $FFT$ spectrum does not depend on the particle position and it closely mirrors the effective shape factor $P(q)$. When the particle reaches the image boundary, instead (panel \ref{fig:1}(c)), a spurious signal is generated, which affects in particular the largest wave-vectors, where the amplitude of the "bulk" signal is lower. This extra contribution appears as a "band" localized around the axis and perpendicular to the image boundary, whose amplitude reach a maximum when the particle is cut in half by the image boundary (panel \ref{fig:1}(d)). If one thinks of the particle displacement as a dynamical process, the temporal persistence of this extra contribution corresponds to the time needed for the particle to completely cross the boundary. In the case of a Brownian particle, this characteristic time can be estimated as $\tau_P \approx \sigma_P^2/D_{t}$, where $D_{t}$ is the particle diffusion coefficient and $\sigma_P$ is the width of its effective shape, which is the largest number between the particle size and the size of the microscope point-spread-function. If a large number of particles is imaged, the boundary contribution is expected to be always present and to fluctuate with the same characteristic correlation time $\tau_P$.

To quantitatively assess this effect, we performed a direct numerical integration of Eq. \ref{gtilde} for the case of a collection of independent Brownian particles. The shape of the particles is described by a Gaussian profile (standard deviation $\sigma_P=0.05$) and the window function $W_0$ is chosen as the characteristic function of the square with unit side length. We adopt time units such that the diffusion coefficient of the particles is $D_{t}=1$ and assume that the axial dynamics can be neglected i.e. that $f(\mathbf{q},\Delta t) \simeq f_{2D}(\mathbf{q},\Delta t)=\exp(-D_{t}q^2\Delta t)$. As a consequence of the spectral leakage, we find that for $q>1/\sigma$ the intermediate scattering functions are no longer described by a simple exponential function and tend to decay with a $q$-independent characteristic time $\tau_P \simeq 2.5$ $10^{-3}$ (Fig. \ref{fig:df}b).

Our simulations indicate that these \textit{dynamic }artifacts can be avoided if one employs the same windowing procedure that is popular in the signal processing community for the removal of spurious \textit{static} signal correlations \cite{Priemer1990,Harris1978}. Windowing consists in multiplying the data, before performing the $FFT$ operation, by a window function, usually a symmetric, bell-shaped profile that smoothly goes to zero at both ends of the sampling interval. In this way, the virtual periodic signal that the FFT algorithm produces by combining an infinite collection of replicas of the original image is no longer discontinuous at the boundaries between tiles. In our case, we find that spatial windowing (Fig. \ref{fig:df}(d)) has a dramatic effect on the reconstructed dynamics (Fig. \ref{fig:df}(e-f)): all the intermediate scattering functions that were previously shown to be corrupted by finite-size artifacts, now display a clean exponential relaxation with the expected relaxation rate $\Gamma(q)=D_{t}q^2$.

The spatial window function chosen in the numerical calculations above and also used in the rest of this article is a Blackman-Harris window function $W_{BN}(x)W_{BN}(y)$, a generalized cosine window function whose 1D version reads \cite{Harris1978}:
\begin{equation}
W_{BN}(x)=\sum^{3}_{j=0} (-1)^j a_{j}\cos \left( \frac{2\pi j x}{L} \right).
\label{wind_funct}
\end{equation}
Here $a_0=0.3635819$, $a_1=0.4891775$, $a_2=0.1365995$, $a_3=0.0106411$. We also tested other options for the window function (Hann and Dolph-Chebyshev \cite{Harris1978}), obtaining equivalent results.  

\section{Spatial windowing in dynamic microscopy experiments}
To assess the validity of the proposed approach in real experiments, we evaluate in this Section the effect of spatial windowing on experimental data acquired with bright-field and confocal microscopy. We will show that spatial windowing of the images before performing the standard DDM analysis drastically reduces the impact of boundary-related artifacts on both the statics and the dynamics.

\subsection{Confocal microscopy}
\label{sec:2}
\begin{figure*}
\resizebox{2\columnwidth}{!}{
\includegraphics{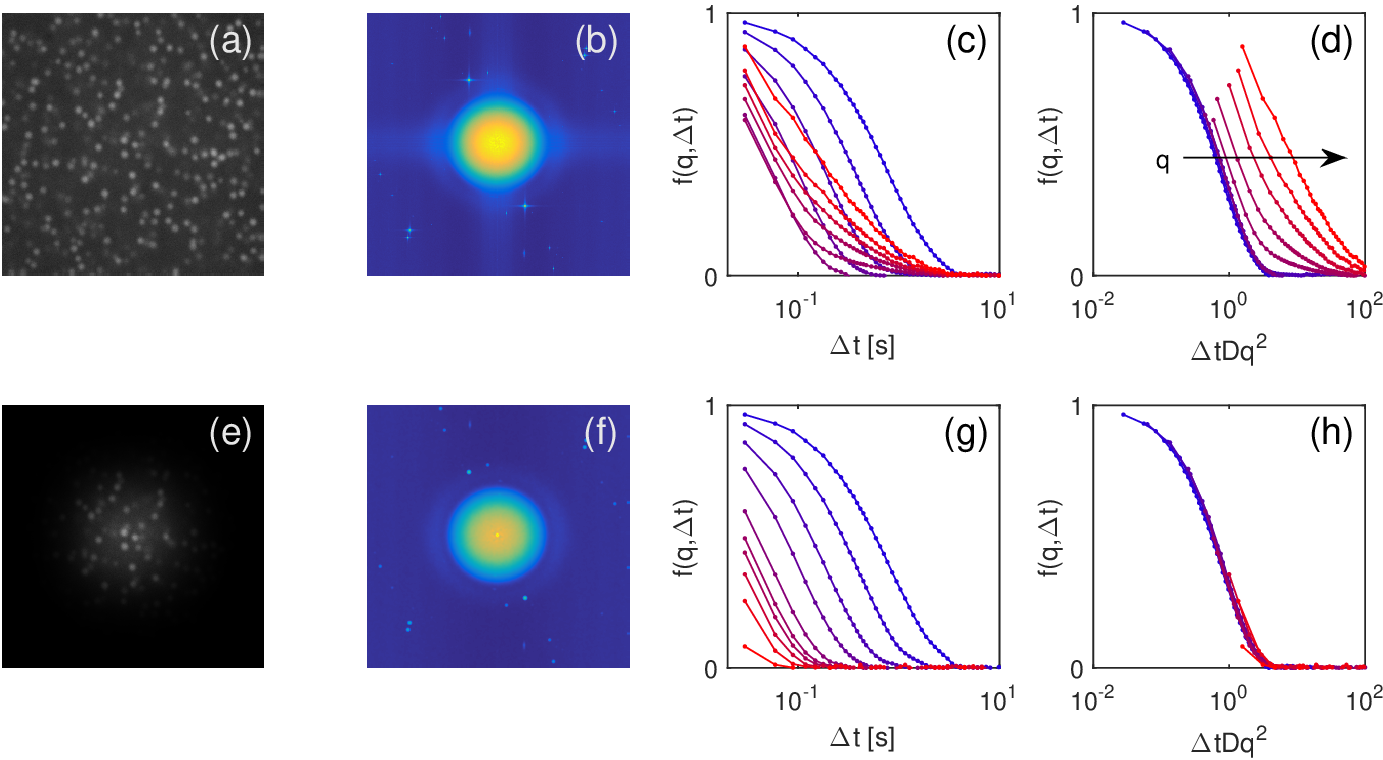}
}
\caption{(a) Representative raw confocal image of a semi-diluted suspension of hard-sphere colloidal particles (see main text for details). 
(b) 2D image structure function $D(\mathbf{q},\Delta t)$ for $\Delta t\simeq 20$ $s$, showing the characteristic "bands" along the axes due to spectral leakage. (c) Azimuthally-averaged image structure functions as a function of the time delay $\Delta t$ for different values of the wave vectors $q$ in the range $2$ $\mu\mathrm{m}^{-1}<q<15$ $\mu\mathrm{m}^{-1}$. (d) Same as in panel (c), but as a function of the reduced delay time $\Delta t D q^2$. The fact that, at large $q$, the curves fail to collapse indicates that the spurious dynamics becomes dominant. (e) Same image shown in panel (a) after spatial windowing with a Blackman-Harris window function (Eq. \ref{wind_funct}). (f) 2D image structure function for $\Delta t\simeq 20$ $s$ as obtained from the windowed image sequence, showing a nice azimuthal symmetry and no bands. The azimuthally-averaged image structure functions are plotted as a function of the time delay $\Delta t$ in panel (g) and of the reduced time delay $\Delta t D q^2$ in panel (h), for the same $q$-values considered in panels (c-d). After windowing, the image structure functions do not show any significant deviation from a purely exponential relaxation with diffusive scaling of the relaxation rate $\Gamma(q)=D_{t}q^2$.}
\label{fig:confocal_1}       
\end{figure*}

The sample a semi-diluted ($0.04$ volume fraction) suspension of sterically stabilized polymethylmethacrylate (PMMA) $0.5$ $\mu m$ fluorescent particles in a density- and refractive index- matching solvent \cite{Lu:2012oa}.
The suspension is imaged by a confocal microscope equipped with a Nipkow disk [Yokogawa], a CCD camera [QIimaging], a 100X oil immersion objective [Leica], and a solid-state laser source [Laserglow].
Image sequences of a single plane from a depth of $20$ $\mu m$ from the lower coverslip are acquired at a frame rate $1/\Delta t_{0}=33.9$ $fr/s$. Image size is 256x256 pixels, with an effective pixel size of $127$ $nm$.

A representative image of the suspension is shown in Fig. \ref{fig:confocal_1}(a). The corresponding two-dimensional image structure function for a large time delay $\Delta t=20$ $s$ (Fig. \ref{fig:confocal_1}(b)) shows marked artifacts, mainly localized along the horizontal and the vertical axis. The impact on the dynamics can be well appreciated from Fig. \ref{fig:confocal_1}(c-d), where the image structure functions obtained from DDM analysis are shown for different values of $q$ in the range $2$ $\mu\mathrm{m}^{-1}<q<15$ $\mu\mathrm{m}^{-1}$. Some of the curves appear non-exponential when plotted as a function of the time delay $\Delta t$ (panel c) and do not collapse on a unique master curve when plotted as a function of $\Delta t D_{t} q^2$ (panel d).

The effectiveness of windowing in amending these effects can be appreciated in Fig. \ref{fig:confocal_1} e-h. A representative 2D structure function obtained for the time delay $\Delta t=20$ $s$ by analyzing windowed images such as the one in Fig. \ref{fig:confocal_1}(e) is shown in (Fig. \ref{fig:confocal_1}(f). It is evident that the expected azimuthal symmetry is recovered. In addition, the temporal dependence of the image structure functions at different $q$s now exhibits the expected exponential decay, with a rate $\Gamma(q) \simeq D_{t}  q^2$ (Fig. \ref{fig:confocal_1}(g-h)).

\begin{figure}
\resizebox{1\columnwidth}{!}{%
  \includegraphics{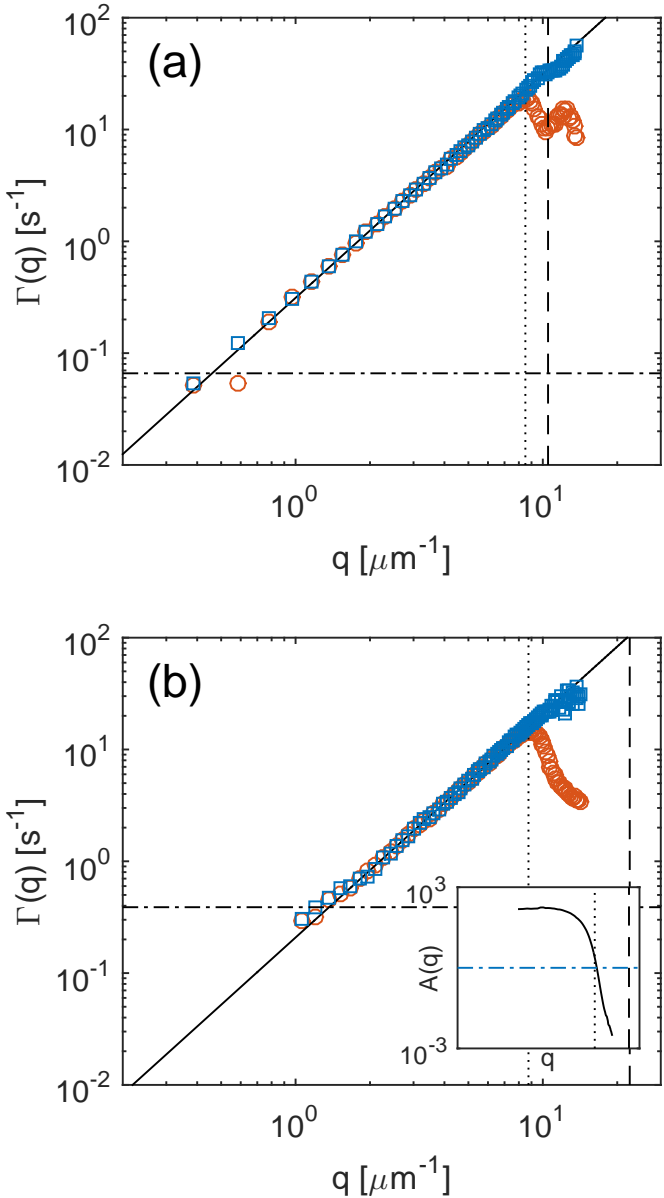}
}
\caption{$q$-dependent correlation rates $\Gamma(q)$ obtained from the fit of the image structure function with (blue squares) and without (orange circles) spatial windowing for the two experiments described in the main text: (a) confocal microscopy of a semi-diluted suspension of $\simeq 500$ nm PMMA particles and (b) bright-field microscopy of a diluted suspension of $\simeq 220$ nm polystyrene particles. In both panels, the horizontal dashed-dotted line represents the inverse of the largest delay time considered in the fit of the image structure functions (roughly corresponding to one tenth of the experiment duration), the vertical dotted line marks the $q$-value for which the unwindowed analysis fails, and the vertical dashed line is drawn in correspondence of position the first minimum of the shape factor $q^* \simeq 4.4934/R$, where $R$ is the particle's radius.
In the inset of panel (b), the amplitude $A(q)$ of the bright-field signal (continuous black line) and of the average noise (blue dashes dotted line) are plotted in arbitrary units over the same $q$ interval as in the main panel. The vertical lines are as in the main panels.}
\label{fig:RATES}       
\end{figure}

To better compare the results of the standard DDM analysis with those obtained by prior windowing of the images, we show in Fig. \ref{fig:RATES}(a) the relaxation rate $\Gamma(q)$ obtained by fitting the image structure functions in Fig. \ref{fig:confocal_1} (c) and (e) with the model $f(q,\Delta t)=\frac{e^{-\Gamma(q) \Delta t}}{\sqrt{1+\gamma\Delta t}}$. In this expression, obtained by assuming an isotropic diffusive dynamics and a Gaussian-Lorentzian model for the confocal point-spread function \cite{Giavazzi:2014sj,Lu:2012oa}, the denominator accounts for the axial dynamics and the $q$-independent rate $\gamma$ is the one associated with the diffusion across the confocal optical section \cite{Giavazzi:2014sj,Lu:2012oa}. 
If we focus only on the horizontal dynamics, the obtained values for $\Gamma(q)$, both in the absence and in the presence of windowing, are compared in Fig. \ref{fig:RATES}(a). In the absence of windowing, a systematic deviation from the expected scaling $\Gamma(q)=D_{t}q^2$ is observed for $q>8.5$ $\mu m^{-1}$, where a sudden drop is observed. On the contrary, windowing allows the reliable reconstruction of the dynamics up to $q \simeq 16$ $\mu m^{-1}$, a limit determined only by the acquisition frame rate that inhibits the access to timescales shorter that about $\Delta t_0$.

Interestingly, the increased wave-vector range made available by the windowing procedure is such that a minimum in the static amplitude is now visible for $q^* \simeq 10.5$ $\mu m^{-1}$. This minimum, corresponding to the dark ring around the central lobe of the Fourier spectrum in Fig.\ref{fig:confocal_1}(f), may be attributed to a zero in the particle's form factor. For a sphere of radius $R^*$ the first zero in the from factor is expected to occur for $q^*=4.4934/R^*$ \cite{Pusey1984}, which provides the estimate $R^*= 4.4934/q^* \simeq $ $0.44$ $\mu m$ for our particles. This value is smaller than the one obtained with the same particles in a previous study \cite{Lu:2012oa}, where a series of measurements were performed for different volume fractions in the range $0.005<\phi<0.4$. In Ref. \cite{Lu:2012oa}, by measuring the diffusion coefficient in a very dilute sample the estimate $R_{H}=0.505$ $\mu m$ was obtained for the particle's hydrodynamic radius. This value was also found to be in good agreement with the size obtained from the Percus-Yevick fit of the static structure factors of the hard spheres.
The observed difference may be attributed to the known fact that for these particles the optical signal is generated by the emission of a fluorescent dye that is physically trapped within the particle itself, in a region that is smaller than the physical size of the particle \cite{Kodger:2017mw}. For this reason, $R^*$ provides an estimate of the size of the fluorescent portion of the particle. It is thus not surprising that $R^*<R_{H}$, also because a) the particles are coated with a non-fluorescent layer of polymer and b) the dye diffuses out of the particle, causing a dye-depleted layer at its surface \cite{Lu:2012oa,Kodger:2017mw}.

\begin{figure}
\resizebox{1\columnwidth}{!}{
\includegraphics{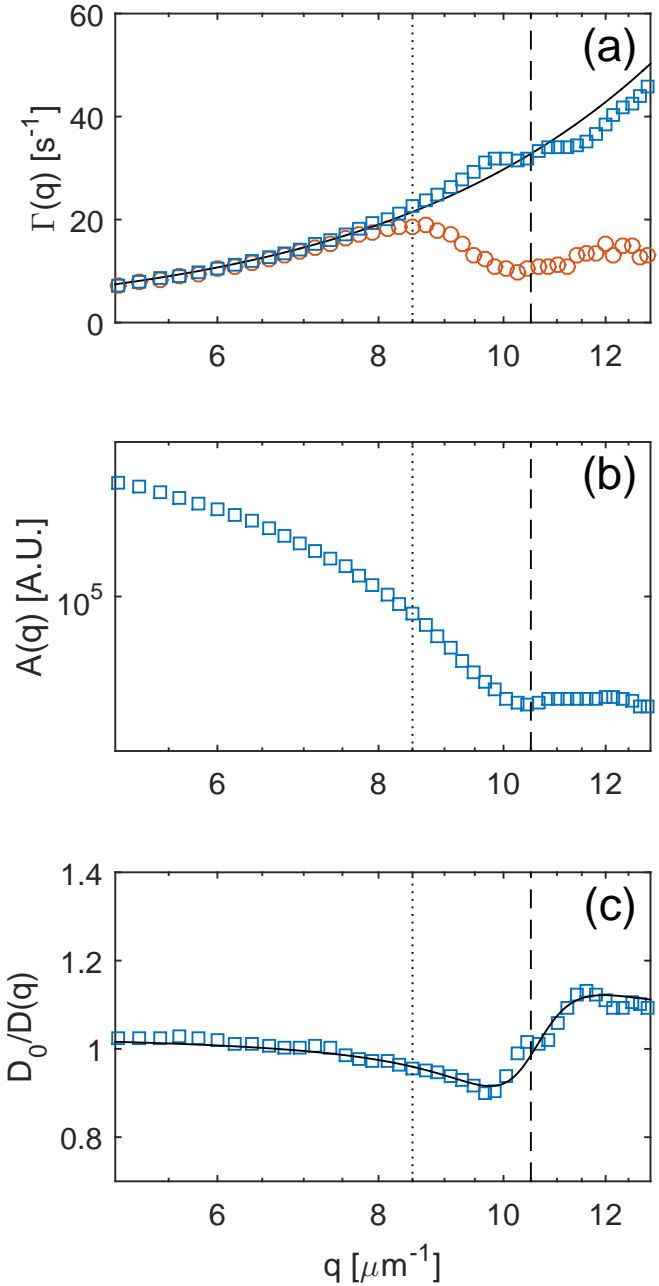}
}
\caption{(a) $q$-dependent decorrelation rates $\Gamma(q)$ obtained from the fit of the image structure function, with (blue squares) and without (orange circles) spatial windowing using the window function in Eq.~\ref{wind_funct}. The continuous line is the best-fit curve to the data with a quadratic function; the vertical dotted line, drawn for $q=8.5$ $\mu m^{-1}$, marks where the curve  $\Gamma(q)$, obtained without windowing, starts deviating significantly from the expected scaling. (b) Amplitude A(q) obtained from the fit of the image structure function after windowing (blue squares), with lines as in (a). (c) Ratio between the effective, $q$-dependent diffusion coefficient $D_t(q) = \Gamma(q) / q^{2} $ and its mean value $D_0$ (blue squares); continuous line is a best-fit to the data with Eq. 33 in Ref. \cite{Pusey1984}. The inflection point in $D_{0}/D_{t}(q)$ occurs for $q^*=10.5$ $\mu$m$^{-1}$, in very good agreement with the position of the first minimum in $A(q)$ (vertical dashed line in all panels). }
\label{fig:confocal_rates}       
\end{figure}

The improved visibility of the minimum in the static amplitude is accompanied by its dynamical counterpart, which brings in additional physics. Careful inspection of the behavior of $\Gamma(q)$ in the vicinity of $q^*$ (Fig. \ref{fig:confocal_rates}(a)) reveals the presence of a characteristic swing on top of the average diffusive scaling $D_{t} q^{2}$, consisting in a slight speed up of the dynamics for $q<q^*$, followed by a slowing down for $q>q^*$. This effect has been predicted and observed in the context of dynamic light scattering \cite{Pusey1984,schope2007} and can be ascribed to the polydispersity of the particles. Let us consider a slightly polydisperse collection of spheres of average radius $\bar{R}$ and polidispersity $\sigma$. For small scattering angles (low $q$), all the particles contribute more or less equally to the scattering signal, which shows a relaxation rate determined by the average diffusion coefficient. For larger $q$, in correspondence of the transferred momentum $q \simeq \frac{4.4934}{\bar{R}(1+\sigma)} \simeq \frac{4.4934}{\bar{R}} (1-\sigma) $ the larger (and slower) particles have a zero in the form factor. As a consequence, at that $q$, they do not contribute anymore to the scattering signal, which thus is dominated by the smaller (and faster) particles. On the contrary, for $q \simeq \frac{4.4934}{\bar{R}}(1+\sigma)$, the main contribution to the dynamics is expected to come from the slower particles since the smaller ones are in the vicinity of a zero in their form factor. The normalized fluctuation $D_0q^2/\Gamma(q)$ is well fitted to the expression given in Eq. 33 in Ref. \cite{Pusey1984} from which a polydispersity $\sigma$ of about $10\%$ can be estimated.
It is important to note that the, although the presence of the "swing" is a strong indication that in our system a distribution of relaxation times is present, its quantitative interpretation must be taken \textit{cum grano salis}. In fact, also according to the previous discussion, the detailed shape and position of this feature are expected to be strongly dependent on the details of the dye distribution within the particle, which is not precisely known. Nevertheless, our findings are compatible with a generic monotonic relationship between the effective radius of the fluorescent portion of the particle and its physical size.

\subsection{Bright-field microscopy}
\label{sec:3}

To test the generality of the proposed approach with respect to the imaging contrast mechanism, we applied the same procedure described in previous paragraph to the case of a very diluted suspension of sub-diffraction colloidal particles imaged with bright-field microscopy. As sample, we chose a suspension of monodisperse polystyrene colloidal particles of nominal radius $R= 230 \pm 10$ $nm$ and volume fraction $\phi \simeq 0.0014$ in a dispersing medium made of water (51.2\% w/w) and glycerol. Bright-field images are collected with a water immersion objective ($40 X$, $NA=1.15$) mounted an inverted microscope (Nikon Eclipse). The microscope is equipped with a fast CMOS camera (Hamamatsu ORCA Flash4 v2, effective pixel size $0.163$ $\mu m$). Sequences of images were acquired with frame rate $1/\Delta t_0 = 777$ $s^{-1}$. DDM analysis was performed both on temporal sequences of raw images and of Blackman-Harris windowed (see Eq. \ref{wind_funct}) images. 

Fitting the temporal dependence of the azimuthally-averaged image structure functions $D(q,\Delta t)$ with a simple exponential decay provides the $q$-dependent relaxation rates $\Gamma(q)$ shown in Fig. \ref{fig:RATES}(b). Fitting of the data for $q\ll 9 $ $\mu m^{-1}$ with $\Gamma(q)=D_{t}q^2$ provides the estimate $D_{t}=0.208 \pm 0.005$ $\mu m^{-1}$. For $q>9 $ $\mu m^{-1}$, we observe that the results obtained without windowing deviate systematically from the expected diffusive scaling $\Gamma(q)=D_{t}q^2$. Such deviation is due to the increasing relevance of the spurious, $q$-independent dynamics of the particles that diffuse in-and-out of the region of interest across its edges. The characteristic rate of the latter process can be roughly estimated as $\tau_P^{-1} \simeq D_{t}/ \sigma_P^2 \simeq 3$ $s^{-1}$, which is compatible with the saturation trend observed for the largest $q$ in Fig. \ref{fig:RATES}(b). We note that for the previous estimate we have used $\sigma_P \simeq\lambda/(2 NA) \simeq 0.26$ $\mu m$, our particle size being below the diffraction limit.

On the contrary, the analysis of the windowed sequence provides consistent results up to $q \simeq 14$ $\mu m^{-1}$, this limit being only set by the signal-to-noise ratio. As it can be appreciated in the inset of Fig. \ref{fig:confocal_rates}(b), the amplitude $A(q)$ is about $400$ smaller than the noise $B(q)$ for $q \simeq 14$ $\mu m^{-1}$. The reliable extraction of quantitative static and dynamic information under this rather unfavorable signal-to-noise ratio is made possible by the use of windowing, which rejects efficiently the finite image-size artifacts. Without windowing the dynamics becomes corrupted as soon as the amplitude of signal falls below the noise level, as spectral leakage effects dominate the signal.

%

%
\section{Conclusion}
We have demonstrated that, in a DDM experiment, particles crossing the boundaries of the images limit and distort the genuine dynamics at high-$q$. The associated $q$-independent dynamic signal leads to a spurious suppression of the relaxation rates measured at large $q$ (Fig. \ref{fig:RATES}). This peculiar feature appears in several DDM-related investigations (e.g. in Refs. ~\cite{Reufer2012,Wulstein:16,Germain:2015gf}), and has thus far not yet been explicitly discussed, nor its origin investigated or explained. In response, we propose a simple solution -- applying a smooth window function to the images before the standard Fourier processing, which despite its conceptual and computational simplicity, significantly enhances the DDM analysis and extending the $q$-range over which meaningful, reliable estimates of the statics and of the dynamics are obtained. However, our solution may have some potential limitations, including overall decrease in signal (typically of about 50\%), due to the reduction of the effective field of view; and the "line broadening" effect, as multiplication of a window function in real space leads to convolution in the Fourier domain \cite{Priemer1990}. Nevertheless, in most practical cases, where both the static amplitude $A(q)$ and the dynamics encoded in $f(q,\Delta t)$ are smooth functions of $q$, this effect should have minimal impact. Consequently, we believe applying a smooth window function as a preprocessing step before Fourier analysis should be an integral part of most DDM implementations, and may also have positive impact in other digital Fourier Microscopy methods \cite{Giavazzi:2014sj}, such as near field scattering or shadowgraphy.

\section*{Acknowledgments}
This work was supported in part by the Italian Ministry of University and Scientific Research (MIUR) (Project RBFR125H0M); Regione Lombardia; the CARIPLO foundation (Project 2016-0998); NASA (NNX13AQ48G); the National Science Foundation (DMR-1310266); and the Harvard Materials Research Science and Engineering Center (DMR-1420570).

\%bibliographystyle{epj}

\end{document}